\documentclass[11pt]{article}
\usepackage{moriond,epsfig}

\def\be{\begin{equation}}
\def\ee{\end{equation}}
\def\bea{\begin{eqnarray}}
\def\eea{\end{eqnarray}}

\begin{document}
\vspace*{4cm}
\title{DIFFRACTIVE DISSOCIATION INTO TWO JETS}

\author{ D. YU. IVANOV${}^{1,2}$ }

\address{${}^1$Institut f\"ur Theoretische Physik, Universit\"at Regensburg, 
\\
D-93040 Regensburg, Germany\\
${}^2$Institute of Mathematics, 630090 Novosibirsk, Russia}

\maketitle\abstracts{
We report a study of
the processes of coherent a pion and a photon
diffraction dissociation into two jets.
The structure of non-factorizable contributions
in these reactions is discussed.
We argue that production of hard dijets by 
real photons can provide direct
evidence for chirality violation in hard processes and the
first measurement of the magnetic susceptibility of the quark condensate.}

\section{Introduction}

The process of a pion diffraction dissociation into a
pair of jets on a nucleon target 
was suggested 
as a probe of the nuclear filtering of pion components
with a small transverse size.
The A-dependence  of the
coherent dijet cross section was first calculated
in \cite{FMS93} and it was argued that the jet
distribution with respect to  the longitudinal momentum fraction has to
follow the quark momentum distribution in the pion and hence provides
a direct measurement
of the pion distribution amplitude. Recent experimental data
by the E791 collaboration \cite{E791a} indeed confirm the strong
A-dependence which is  a signature for color transparency. 
Moreover, the jet longitudinal
momentum fraction distribution turns out to be consistent with the
$\sim z^2(1-z)^2$ shape corresponding to the asymptotic pion distribution
amplitude. After these first successes, one naturally asks whether the
QCD description of coherent dijet production can be made fully quantitative.
First we discuss factorization and
concentrate on a pion dissociation process, 
then
we will consider production of dijets initiated by a real photon,
a process which
is sensitive to the chiral--odd properties of QCD vacuum.
The results reported here have been obtained in collaboration with
V. Braun, S. Gottwald, A. Sch\"afer and L. Szymanowski
\cite{BISS01,Braun:2002en}.

\section{Pion dissociation}

\begin{figure}
\centerline{\psfig{figure=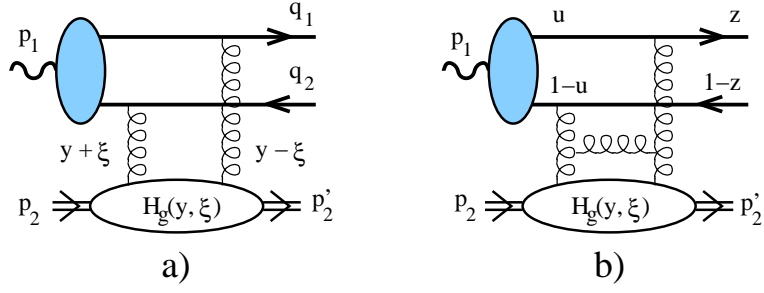,height=1.5in}}
\caption{Sample diagrams for the hard dijet production, see text.
\label{fig:1}}
\end{figure}

The kinematics of the process and the notations for momenta 
is shown in Fig.~\ref{fig:1}.
The momenta of the incoming particles  are $p_1$ and $p_2$,
$z$ is the longitudinal momentum fraction ($\bar z\equiv 1-z$)
and $q_\perp$ the transverse momentum of the quark
jet.
We consider the forward limit,
when transverse momenta of the jets compensate each other.
In this kinematics the invariant mass of the produced $q\bar q$ pair is
equal to
$
M^2={ q_{\perp}^2}/{ z\bar z}
$,
and the momentum of the outgoing nucleon $p_2^\prime=p_2(1-\xi)/(1+\xi)$,
where
$\xi= M^2/(2s-M^2) \simeq M^2/2s$, $s=(p_1+p_2)^2$.
%

The possibility to constrain the pion distribution amplitude
$\phi_\pi (u)$ in the dijet diffractive dissociation
experiment assumes that
the amplitude
of this process can be calculated in the collinear
approximation as
suggested by  Fig.~\ref{fig:1}:
\begin{equation}
{\it M}
=\int\limits^1_0 du \int\limits^1_{-1} dy
\,\phi_\pi
(u)
\,T^g_H (u , y)\,{\cal H}_g(y,\xi )\,.
\label{factor}
\end{equation}
Here ${\cal H}_g(y,\xi)$ is the generalized gluon
distribution in the target nucleon
\cite{Ji97a};
variable $-1<y<1$ parametrizes the
momentum fractions of the emitted and the absorbed gluons.
$T_H^g(u, y)$ is the hard scattering amplitude involving at
least one hard gluon exchange.

There are two important regions in the integral  (\ref{factor}), see
\cite{BISS01} for more details. At
$u\to 0,1$
\begin{equation}
 {\it  M}\Big|_{\rm end-points}
  \sim
i\,  z \bar z \,  \int\limits^1_{u_{\rm  min}} du\,
\frac{\phi_\pi(u)}{u^2}{\cal H}_g(\xi,\xi) \,.
\label{end}
\end{equation}
Since $\phi_\pi(u)\sim u$ at $u\to 0$, the integral over $u$
diverges logarithmically.
Remarkably, the integral containing the pion distribution
amplitude does not involve any $z$-dependence.
Therefore, the longitudinal momentum distribution of the jets in the
nonfactorizable contribution is calculable and, as it turns out,
has the shape of the asymptotic pion distribution amplitude
$\phi_\pi^{\rm as}(z) = 6z\bar z$.
The appearance of the end point divergence
is due to pinching of the $y$ contour at the point $y=\xi$
in case that the variable $u$
is close to the end--points.
One can trace \cite{BISS01} that this pinching occurs
between soft gluon
interactions in the initial and in the final state, and is related with
the existence of the unitarity cuts of the amplitude in different,
$s$ and $M^2$, channels.
The other important integration region in Eq. (\ref{factor}) is the one
when the longitudinal momentum fraction carried by the quark is close
(for high energies) to that of the quark jet in the final state
\begin{equation}
{\it M}\Big|_{\xi\ll |u- z|\ll 1}
\sim 4 i \phi_\pi(z)\!\int\limits_\xi^1
 \!\frac{dy}{y+\xi}\, {\cal H}_g(y,\xi)\,.
\label{z=z'}
\end{equation}
This logarithmic integral is nothing but
the usual energy logarithm that accompanies each extra gluon in the
gluon ladder. Its appearance is due to the fact the hard gluon
which supplies jets by the high transverse momentum
can be emitted in a broad rapidity interval and
is not constrained to the pion fragmentation region.
The integral on the r.h.s. of Eq. (\ref{z=z'}) can be identified with the
unintegrated generalized gluon distribution.
Therefore, in this region hard
gluon exchange can be viewed as a large transverse momentum part of the
gluon distribution in the proton, cf. \cite{NSS99}.
This contribution is proportional to
the pion distribution amplitude $\phi_\pi (z)$
whereas the end-point contribution (\ref{end}) imitates the shape of $\phi^{\rm{as}}_\pi
(z)$. This implies, in a contradiction to \cite{FMS93}, that 
jets longitudinal momentum distribution 
does not proportional to $|\phi_\pi (z)|^2$.

\section{Photon dissociation}

The wave function of a real photon contains both the
perturbative chiral-even (CE) contribution of
the quark-antiquark pair with opposite helicities, and the nonperturbative
chiral-odd (CO) contribution with quarks having the same helicity and which
is due to the chiral symmetry breaking. It is proportional to fundamental
parameters of QCD vacuum, quark condensate $\langle \bar q q \rangle$
and magnetic susceptibility $\chi$.
The perturbative CE contribution is
singular $\sim 1/|{\bf r}|$ at small transverse distances ${\bf r}$.
The nonperturbative CO contribution is regular at small transverse
separations
and can be parametrized by the photon
distribution
amplitude $\phi_\gamma(u,\mu)$ \cite{BBK89}
\begin{equation}
\langle 0 |\bar q(0) \sigma_{\alpha\beta} q(x)
   | \gamma^{(\lambda)}(q)\rangle = i \,e_q\, \chi\, \langle \bar q q
\rangle
 \left( e^{(\lambda)}_\alpha q_\beta-  e^{(\lambda)}_\beta q_\alpha\right)
 \int\limits_0^1 \!du\, e^{-iu(qx)}\, \phi_\gamma(u,\mu)\,.
\label{phigamma}
\end{equation}
$\phi_\gamma(u, \mu \ge 1~\mbox{\rm GeV})$
is believed to be not far from the asymptotic form $
   \phi_\gamma^{\rm as}(u) = 6 u (1-u)$.
$\chi$ was estimated using the vector dominance
approximation and QCD sum rules~\cite{chiold,chi}:
$\chi \langle \bar q q \rangle \simeq 40-70~\mbox{\rm  MeV}$.
However, any direct experimental evidence on both $\chi$ and
$\phi_\gamma(u)$
is absent.
This structure can be studied in experiments
similar to the studies of coherent dijets in
pion dissociation by the E791 collaboration \cite{E791a}.

Since the CE and CO contributions lead to final states with
different helicity, they do not interfere and the dijet
cross section is given by the incoherent sum, for the linearly polarized
photon
\begin{equation}
\frac{d\sigma_{\gamma\to 2\,{\rm jets}}}{d\phi
d q_\perp^2 dt dz}{\Bigg|}_{t=0}
\ =\ \sum_q e^2_q \frac{\alpha_{EM} \alpha_s^2 \pi^2
(1+\xi )^2
}{4\pi N_c q_\perp^6}
\left[(1-4z\bar z \cos^2\!{\phi})
|{\cal J}_{CE}|^2+\frac{\pi^2 \alpha_s^2 \chi^2 \langle
\bar q q \rangle^2 }{N_c^2q^2_\perp}|{\cal J}_{CO}|^2\right], 
\label{c.sect.}
\end{equation}
where $\phi$ is the azimuthal angle between the jet direction and
the photon polarization $(e^{(\lambda)}\cdot q_\perp) \sim \cos\phi$,
${\cal J}_{CE}$ and ${\cal J}_{CO}$ are the CE and CO amplitudes
respectively.
Note that the CE contribution is $\sim 1/q_\perp^6$ \cite{FMS93} and the
CO contribution is suppressed by one extra
power of $q_\perp^2$ which follows from twist counting.
The different $\phi$ dependence can be traced to the fact that
the $q\bar q$ pair is produced in a state with orbital angular momentum
$L_z=0$ and $L_z=\pm 1$ for the CO and CE contributions, respectively.
The CE contribution originates from the region of large momenta flowing
through
the photon vertex. 
To leading order (LO) in the strong coupling $\alpha_s = \alpha_s(q_\perp)$
the amplitude is given by the sum of Feynman diagrams of the type
shown in Fig.~\ref{fig:1}a 
\begin{equation}
{\cal J}_{CE}= i \xi {\cal H}_{g}^\prime (\xi,\xi)
+\frac{i \alpha_s N_c}{\pi}\int^1_\xi\frac{dy }{y+\xi}
{\cal H}_{g} (y,\xi)\,,
\label{even}
\end{equation}
where ${\cal H}_{g}^\prime (\xi,\xi)=d{\cal H}_{g} (y,\xi)/dy|_{y=\xi}$.
The second term in Eq.~(\ref{even}) originates from the diagrams with
additional gluon exchange between the
$t$-channel gluons, see Fig.~\ref{fig:1}b, it corresponds to the leading
at
large energies
(enhanced by
$\log\xi$) NLO
contribution.
Since ${\cal H}_{g} (y,\xi)\sim
G(y)$ at $y\gg \xi$, and as the factor
$\alpha_s N_c/(\pi y)$ is nothing but the low-$y$ limit of the
DGLAP gluon splitting function,  the integral in Eq.~(\ref{even}) can be
identified to logarithmic accuracy with the unintegrated gluon distribution
$f(\xi,q^2)=\partial G(\xi,q^2)/\partial \ln q^2$. 

For the nonperturbative CO contribution
the large momenta are not allowed in the
photon vertex and the factorization formula contains a
convolution with the photon distribution amplitude.
In this case an additional hard gluon exchange is mandatory and the
diagram in Fig.~\ref{fig:1}b presents one example of the existing 31 LO
contributions. Similar to the pion case
we found that the result for the amplitude may be
approximated well by the sum of two contributions in analogy to
Eqs.~(\ref{end}) and (\ref{z=z'}).
The origin of the end--point divergence
is the same as in a pion dissociation.
Assuming that the photon distribution amplitude
is close to the asymptotic form,
we obtain ${\cal J}_{CO}\sim z(1-z)$ for both integration regions, up to
small
corrections. The presence of
nonfactorizable contribution does not have, therefore,
any significant effect on the jet distribution but mainly influences the
normalization.

In the numerical calculation performed for  HERA kinematics we
have introduced an infrared cutoff
$u_{\rm  min} = \mu^2_{\rm
IR}/q_\perp^2$ to regularize the nonfactorizable contribution,
$\mu_{\rm IR}=500$~MeV.
We found that the nonperturbative CO contribution
integrated over $\phi$, $z$ and
$t$
is of the order
of
\begin{equation}
 \frac{d\sigma_{CO}}{d\sigma_{CE}} \simeq
   (7\pm 2~\mbox{\rm GeV})^2\cdot \frac{\alpha_s(q_\perp)^2}{q_\perp^2}\,
   \left(\frac{\chi\langle \bar q q\rangle}{50~\mbox{\small \rm
MeV}}\right)^2.
\label{COscale}
\end{equation}
For $q_\perp > 4$~GeV the cross section is dominated by the perturbative CE
contribution, for smaller transverse momenta the
dijet cross
section is saturated by the CO contribution.
The transition between the two different regimes
is seen very clearly from
the dependence of the cross section on the dijet longitudinal momentum
fraction a
the azimuthal angle.
At $q_\perp >4 $~GeV the $z$-distribution is almost flat,
while
the $\phi$ distribution is almost purely $\sim 1-\cos^2\phi$. In contrast to
this
at $q_\perp < 4$~GeV the $z$-distribution is comparable with  $\sim
z^2(1-z)^2$
while the $\phi$-distribution becomes flat.
Our main result is that the nonperturbative CO contribution is large in the
region
of intermediate $q_\perp\sim 2-4$~GeV and can be clearly separated from the
perturbative contribution by a different $z$- and $\phi$-dependence.
Observation of the CO contribution would be the first clear
evidence for the chirality violation in hard processes and also provide the
first direct measurement of the magnetic susceptibility of the quark
condensate.

\section*{Acknowledgments}
The author thanks the organizers of the QCD Moriond 2003 for the kind
invitation. This work is supported by DFG and BMBF (06OR984) and in part
by INTAS 00-00679.

\end{document}